# Model-based cellular kinetic analysis of SARS-CoV-2 infection: different immune response modes and treatment strategies

Zhengqing Zhou[1*], Ziheng Zhao[2*], Shuyu Shi[3*], Jianghua Wu[4*], Dianjie Li[1], Jianwei Li[1], Jingpeng Zhang[1], Ke Gui[2], Yu Zhang[2], Heng Mei[4#], Yu Hu[4#], Qi Ouyang[1#], and Fangting Li[1#]

[1] School of Physics, Center for Quantitative Biology, Peking University, Beijing 100871, China
[2] Department of Immunology, School of Basic Medical Sciences, NHC Key Laboratory of Medical Immunology, Peking University, Beijing 100191, China.
[3] Peking University Third Hospital, Peking University, Beijing 100191, China.
[4] Institute of Hematology, Union Hospital, Tongji Medical College, Huazhong University of Science and Technology, Wuhan, 430022, China

* These authors contribute equally to this work
# To whom correspondence should be addressed. dr_huyu@126.com, hmei@hust.edu.cn, qi@pku.edu.cn, lft@pku.edu.cn

## Abstract

Increasing number in global COVID-19 cases demands for mathematical model to analyze the interaction between the virus dynamics and the response of innate and adaptive immunity. Here, based on the assumption of a weak and delayed response of the innate and adaptive immunity in SARS-CoV-2 infection, we constructed a mathematical model to describe the dynamic processes of immune system. Integrating theoretical results with clinical COVID-19 patients' data, we classified the COVID-19 development processes into three typical modes of immune responses, correlated with the clinical classification of mild & moderate, severe and critical patients. We found that the immune efficacy (the ability of host to clear virus and kill infected cells) and the lymphocyte supply (the abundance and pool of naïve T and B cell) play important roles in the dynamic process and determine the clinical outcome, especially for the severe and critical patients. Furthermore, we put forward possible treatment strategies for the three typical modes of immune response. We hope our results can help to understand the dynamical mechanism of the immune response against SARS-CoV-2 infection, and to be useful for the treatment strategies and vaccine design.

## Introduction

Coronavirus disease 2019 (COVID-19) caused by the coronavirus SARS-CoV-2 has spread globally, having a huge impact on global politics, economy and society. Compared to other viral infectious diseases, such as influenza, severe acute respiratory syndrome (SARS), Middle East respiratory syndrome (MERS), and acquired immune deficiency syndrome (AIDS), COVID-19 exhibits multiscale different characteristics. Epidemiologically, COVID-19 has a relatively longer incubation period (~5.8 days) with a number of asymptomatic patients, which intensifies the difficulty of management and prevention (*1*). For the within-host viral infection, about 80% COVID-19 patients exhibit mild symptoms and recover within 3~4 weeks after regular treatments (*2*). The severe and critical COVID-19 patients (~20%) are related to lymphopenia, high neutrophil counts and cytokine release syndrome (CRS) or cytokine storm, characterized by elevated inflammatory cytokines levels like IL-6. CRS and succeeding comorbidities usually cause bad clinical outcome and even death, although the overall intensity of the cytokine storm in COVID-19 patients is milder than SARS patients (*3*).

More studies and evidences show that the SARS-CoV-2 virus, compared to SARS-CoV and MERS-CoV, in the early infection period tends to induce less effective anti-viral innate immune responses with a delayed or lower type-I interferon (IFN) response and lower HLA-II expression level (*4, 5*). Furthermore, marked lymphopenia and impaired humoral immunity with the loss of germinal centers (*6*) suggest that the weak adaptive immune response could contribute to damped clearance of virus and thus chronic infection.

Though the knowledge and clinical data of COVID-19 is increasing, a systemic view of immunologic response in SARS-CoV-2 infection remains necessary. Due to the variance in immune status and response dynamic processes among patients, it is hard to make up an effective therapeutic schedule, for example, the effects of remdesivir, IFN-γ and antibodies remain controversial (*7-13*).

Here, we investigated the immune response against SARS-CoV-2 infection by analyzing the longitudinal hemogram data of 194 patients from Wuhan Union Hospital and also by mathematical modeling the within-host immune dynamics. We constructed within-host virus-immune interaction network, together with a mathematical model depicting the dynamic processes and the response of the innate immunity and adaptive immunity against SARS-CoV-2 infection. We simulated and classified the different modes of patient's immune responses, which correspond to the longitudinal data of COVID-19 patients, and we propose the possible treatment strategies to improve the immune status of COVID-19 patients.

## Immune network and model, immune efficacy and T cell supply

The human immune system is a complex defense system involving dozens of different cell types and hundreds of interacting molecular pathways, protecting human body from dangerous pathogens. A mathematical model, which describes the complex interactions in immune system and the demographic differences in patient's health status, should help us to understand the underlying immune response, to classify the patients based on their immune response processes, and provide possible definitive treatment strategies.

We analyzed 194 COVID-19 patients' longitudinal data from Union Hospital of Tongji Medical College in Huazhong University of Science and Technology (Wuhan, P. R. China), including the hemogram, the serum cytokine profile and treatments at different time points, as well as their clinical classification and outcome. We found that among mild, severe and critical COVID-19 patients, the time sequences of counts of white blood cell, neutrophil and lymphocyte, together with IL-6 levels in peripheral blood behave significant different, indicating the severity of patients (Fig. 2B). In critical patients, the elevated neutrophil counts and myeloid-derived suppressor cell (MDSC) fraction are significantly greater than the mild and severe patients, which could be accredited to the chronic viral and bacterial co-infection (Fig. S5B).

Based on the recent evidences and clinical data about COVID-19, we put forward a possible immune mechanism during SARS-CoV-2 infection: a slow innate immune response in the early infection stage leading to extensive tissue damage and inflammation, and a weak adaptive immune response in later stage resulting in the chronic infection and bad clinical outcome.

We simplified both innate and adaptive immune response processes within-host caused by the SARS-CoV-2 infection, and constructed a virus-immune interaction network with the viral infection module, innate immunity, cellular immunity, humoral immunity, and immunosuppression modules (Fig. 1). During the SARS-CoV-2 infection, infected lung epithelial cells recruit innate immune cells through chemokine secretion, including neutrophils (Neut), macrophages (MΦ), dendritic cells (DC), and natural killer cells (NK). These cells serve as the initial immune defense in the virus-immune interaction network against the virus, and secrete the inflammatory cytokines like IL-6 and TNF-α. The DC and MΦ engulf and process SARS-CoV-2 specific antigens they encounter, and then they work as antigen-presenting cells (APC) to activate naïve T and B cells. The activated CD4+ and CD8+ T cells proliferate and differentiate into effective helper T cells (Th) and cytotoxic T cells (CTL), then travel to the airway and lung fighting against the pathogens. Meanwhile, the germinal centers in lymph nodes form around the pathogen-loaded dendritic cells, where naïve B cells go through affinity maturation with clonal selection and differentiate into plasma cells that produce immunoglobulin (Ig) to clear the virus. During this process, regulatory T cells (Treg) are also activated to prevent the over activation of immune system and turn off the immune response when the virus is cleared.

For the sake of simplicity and clarity of the model, we made the following main simplifications and assumptions. (1) We focus on the host immune response in lung and nearby draining lymph nodes (lung area). (2) IL-6 is selected as the key indicator of inflammation. (3) To compare with clinical data, we add the antiviral drug term (parameter $\alpha$ in Eq.2) in our model. (4) We discuss the primary virus infection and immune response, and ignore the process and function of SARS-CoV-2-specific memory T and B cells. (5) The cytokine level in lung is estimated to be 10 times of the peripheral blood cytokines, providing a reference for comparison between modeling results and clinical data. More details about the interactions among virus, immune cells and cytokines can be found in Table 1 of Supplemental Material (SM), and our other assumptions are listed in section 1.2 of SM.

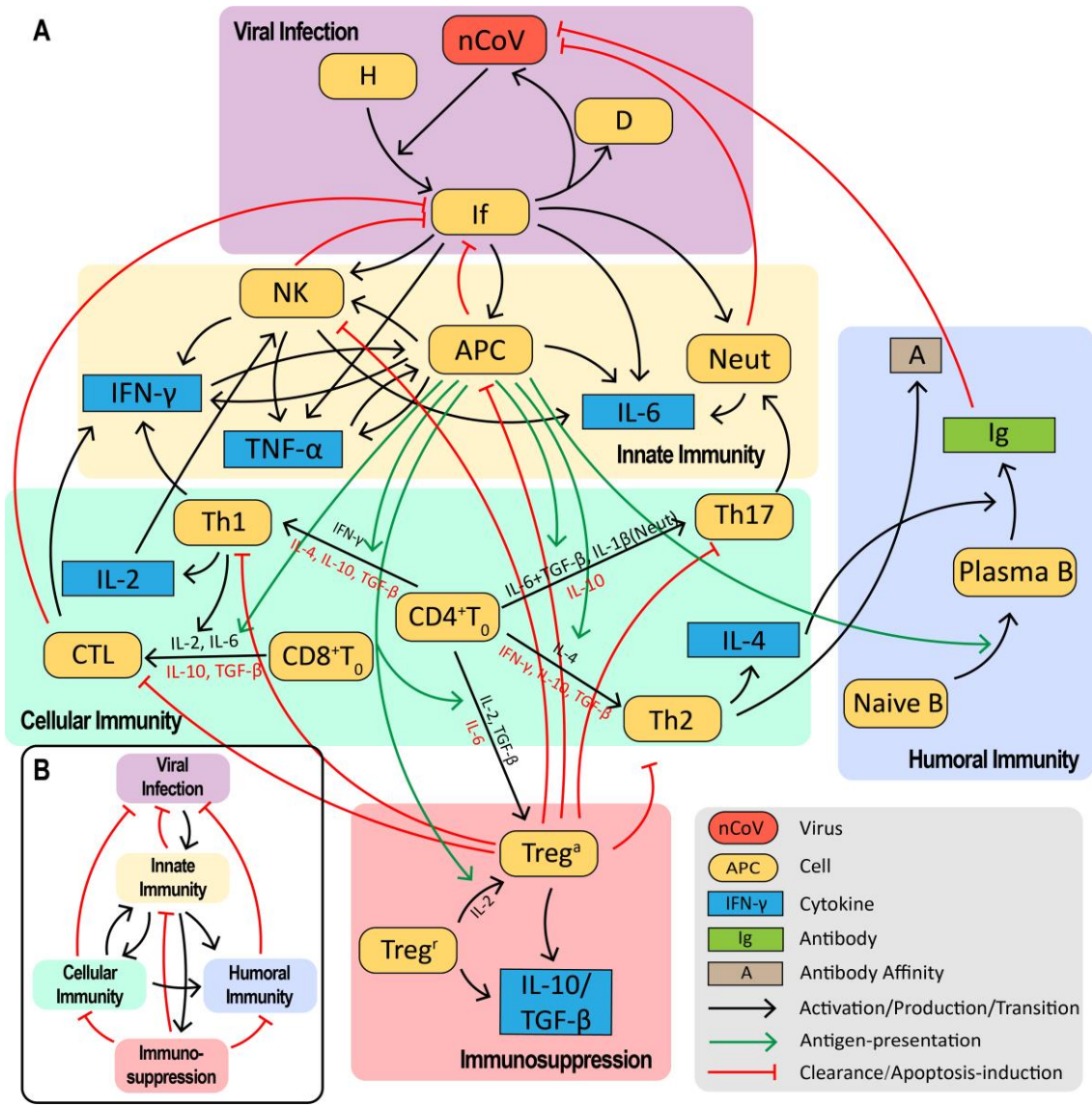

Fig. 1.The immune system response network against SARS-CoV-2 infection. (**A**) The network can be divided into five modules: viral infection, innate immunity, cellular immunity, humoral immunity and immunosuppression modules, each part including complex and nonlinear interactions among immune cells and cytokines. (**B**) The overall interactions among the five modules.

Then we built a 24-variable ODE model to depict the dynamical process of immune response against SARS-CoV-2 infection. Clinical data and results, together with the evidence on the lower IFN-I response (*4, 5*), provide a reference for the construction and parameterization of the model. In lung area, we define [*nCoV*] as the concentration of free viral load, [*H*] and [*If*] denote respectively the concentrations of healthy pulmonary epithelial cell and productively infected cell. The concentrations of neutrophils, APC (DC and MΦ), natural killer cells and cytotoxic T lymphocytes (CTL) are denoted as [*Neut*], [*APC*], [*NK*] and [*CTL*] respectively. [*Ig*] is the concentration of antibodies. The viral load and lymphocytes are in the unit of $10^6$/mL, the unit of cytokines is pg/mL, and the unit of antibodies is μg/mL.

In the viral infection module, we have the following equations to depict how the SARS-CoV-2 virus infects the lung epithelial cells.

$$\frac{d[nCoV]}{dt} = \gamma N_1 d_{If}[If] - (k_1^{clear}[Neut] + k_2^{clear} \cdot A \cdot [Ig])[nCoV] \quad (1)$$

$$\frac{d[If]}{dt} = \alpha k_{infect}[nCoV][H] - \{k_1^{kill}[APC] + k_2^{kill}[NK] + k_3^{kill}[CTL]\}[If] - d_{If}[If] \quad (2)$$

$$\frac{d[H]}{dt} = r_H - \alpha k_{infect}[nCoV][H] - d_H[H] \quad (3)$$

The viral dynamics is described in Eq.1. The virion particles are produced from infected cells at the rate of $\gamma N_1 d_{If}[If]$, where $\gamma$ stands for a conversion factor between particle count and extracellular particle number density, $N_1$ is the burst size of SARS-CoV-2 virus, $d_{If}$ is the death rate of infected cells that release new virions. The viruses are cleared by neutrophils [Neut] and antibodies [Ig] at the rate of $(k_1^{clear}[Neut] + k_2^{clear} \cdot A \cdot [Ig])[nCoV]$. In Eq.2, the epithelial cells are infected by free virions at rate $\alpha k_{infect}[nCoV][H]$, where $\alpha$ stands for the effect of antiviral drugs, $k_{infect}$ is the infection rate of virus. The infected cells are killed by APC (DC and MΦ), natural killer cells and CTL at the rate $\{k_1^{kill}[APC] + k_2^{kill}[NK] + k_3^{kill}[CTL]\}[If]$. The infected cells die at rate $d_{If}[If]$ and release new free virion particles. In Eq.3, the healthy lung epithelial cells regenerate at the rate $r_H$ and undergo normal apoptosis at rate $d_H[H]$.

We define $e^{kill}(t) \equiv k_1^{kill}[APC] + k_2^{kill}[NK] + k_3^{kill}[CTL] + d_{If}$ and $e^{clear}(t) \equiv k_1^{clear}[Neut] + k_2^{clear} \cdot A \cdot [Ig]$. Under adiabatic approximation of change of the infected cells, we have $\frac{d[nCoV]}{dt} = e^{clear}(R_0 - 1)[nCoV]$, where the reproductive ratio is defined as $R_0 \equiv \frac{\alpha \gamma N_1 d_{If} k_{infect}[H]}{e^{clear} \cdot e^{kill}}$. To represent the ability of immune system to clear the virus and kill the infected cells, we define the host immune efficacy at time point $t$ as $e(t) \equiv e^{clear}(t) e^{kill}(t)$. Thus, at time point $t$, if $R_0(t) < 1$ and $e(t) > \alpha \gamma N_1 d_{If} k_{infect}[H]$, the population of virus will decrease; otherwise the population of virus will increase.

When the naïve CD8+T cells meet and interact with antigen-loaded APCs in the lymph nodes, they are activated, proliferate and differentiate to CTL, then move to the lung area to fight against the virus. We use a logistic term to describe the homeostasis of naïve CD8+ T cell supply with capacity $K_{CD8}$ and growth rate $r_{CD8}$, $k_{CTL}$ represents the activation rate to CTL from the naïve CD8+ T cell by APC interaction. Thus, we wrote the dynamics of naïve CD8+T cell as:

$$\frac{d[CD8^+T_0]}{dt} = r_{CD8}[CD8^+T_0]\left(1-\frac{[CD8^+T_0]}{K_{CD8}}\right) - k_{CTL}\frac{[APC]^5}{K_A^5+[APC]^5}[CD8^+T_0] \quad (4)$$

Steady state solution gives out $[CD8^+T_0] = \left(1 - \frac{k_{CTL}}{r_{CD8}} \cdot \frac{[APC]^5}{K_A^5+[APC]^5}\right)K_{CD8}$. We define the host supply ability of naïve CD8+T cell as $s_{CD8} \equiv \frac{r_{CD8}}{k_{CTL}}\frac{K_A^5+[APC]^5}{[APC]^5}$, which indicates the ability of CD8+ T cell supply to produce more CTL. If $s_{CD8} > 1$, there will have sufficient naïve CD8+T cell. Once chronic infection takes place, the patients with $s_{CD8} < 1$ might experience CD8+ T cell exhaustion. When $[APC] > K_A$, $s_{CD8} \to \frac{r_{CD8}}{k_{CTL}}$. Thus, the host supply ability $s_{CD8}$ determine whether the CD8+ T cell pool can supply more CTL cells in the 'killing' infected cell process, which works as $k_3^{kill}[CTL]$ term in the $e^{kill}(t)$ and $e(t)$. Similar supply analysis and results can be applied to CD4+ T cells.

More details concerning immune cell and cytokine dynamics can be found in the section 2 of SM.

Furthermore, we also utilized the model to simulate other viral infectious diseases, including influenza and severe acute respiratory syndrome (SARS). The differences in virus infection rate, burst size and activation strength of the immune system lead to significant differences in epidemical features among influenza, SARS and SARS-CoV-2, especially the incidence rate, incubation period and critical rate. Detailed parameter settings and results are shown in the section 3 of SM.

# Modeling typical modes of immune response against SARS-CoV-2

Demographically, the immune system's capability and response vary individually, susceptible to age, fitness and gender differences (*14, 15*). In Fig. 2A, we adopted Latin hypercube sampling method (*16*) to search across the parameter space and identified three typical modes of the immune responses against SARS-CoV-2 infection, where they differ in the extent of tissue damage, final state viral load and immune response, in particular the levels of host immune efficacy $e$ and supply ability $s_{CD8}$. We also defined asymptomatic patients by underactive immune and inflammatory response. Full trajectory of the three immune modes and asymptomatic patients can be found in Fig. S2A and their definitions in Table S2 of SM. Parameter sample range and initial value choice can be found in section 7 of SM. Sampled curves are aligned at [nCoV]=$10^6$/mL.

During the early stage of infection, the post admission days (PAD) 0~7, a faster and stronger innate immunity is evoked in the mode 1 patients, protecting lung tissue from viral damage. For the mode 2 and 3 patients, delayed and weaker innate immunity brings about more extensive damages with higher viral load, manifested in Fig. S2C and D of SM. During the middle stage (PAD 7~14), the adaptive immunity of mode 1 and 2 patients is built up successfully, providing the patients with strong immune efficacy, averaged at $e_{max} > 12$, which is enough to clear the virus and kill the infected cells. For the mode 2 patients, the accumulated tissue damage at early stage over-activates the innate immunity, especially neutrophils and monocytes and thus causes temporary cytokine storm. The mode 3 patients' immune response stays low at about $e = 3.5$, possibly attributed to limited antibody production and inadequate CD8+ T cell supply. During the late stage (PAD 14+), the mode 1 and 2 patients recover from the infection, however the mode 3 patients experience chronic infection and 23% end up with naïve CD8+T cell pool insufficiency (denoted as insufficient $s_{CD8}$). Time course of $1/s_{CD8}$ and $e$ of the three modes are shown in Fig. 2C and Fig. 2D.

In comparison to mode 2, mode 3 is more likely to appear in older patients with underlying diseases, including but not limited to hypertension and diabetes mellitus, for their fragile immune systems. This view is also supported by recent study on the association between adaptive immunity and age (*18*). In addition to these, in mode 3, long-termed damage of lung epithelial cells and inadequate immune efficacy could possibly induce secondary bacterial or fungal infection, discussed in section 7 of SM, thus lead to overactive inflammatory response and cytokine storm. Therefore, patients with mode 3 response are more likely to be critical and should be paid with extra attention.

Figure 2.

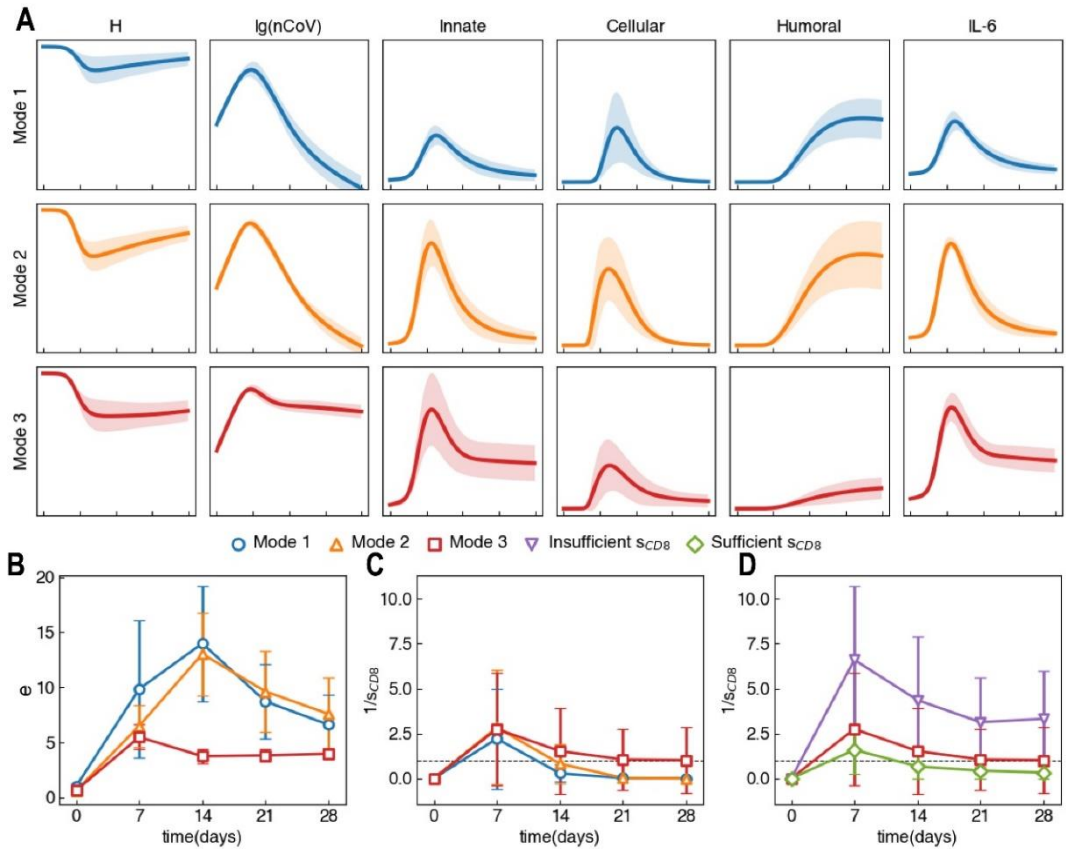

Fig. 2. Sampling result of host immune response against SARS-CoV-2 infection. (A) Schematic illustration of three typical modes of immune response, the sampled curves are aligned at [nCoV]=$10^6$/mL. Innate immunity is defined by the summation of APC and NK density, cellular immunity is defined by CTL density and humoral immunity is defined by antibody level. (B) Time course of immune efficacy $e$ of the three immune modes. During early stage, mode 1 has higher $e$ compared to mode 2 and 3, leading to less extensive tissue damage and thus milder level of cytokine storm. At around day 14, $e$ value of mode 1 and 2 rises to the peak, corresponding to the fully activation of immune system. Meanwhile, mode 3 patients do not show the peak, suggesting underactive adaptive immunity leads to chronic infection. (C) Time course of inverted CD8+ T cell supply $1/s_{CD8}$ of mode 1, 2 and 3. Both mode 1 and 2 pass $s_{CD8} < 1$ temporarily, corresponding to the maximum activation from innate immunity to adaptive immunity, while mode 3 patients show a durative $s_{CD8} < 1$, leading to insufficient CD8+ T cell supply. (D) Mode 3 patients are further divided in to naïve CD8+T cell insufficient ones (insufficient $s_{CD8}$) and sufficient ones (sufficient $s_{CD8}$).

## Classification of patients and relevant treatment strategies

To understand the dynamical features and processes of COVID-19 patients' immune response, we investigated the longitudinal data of hemogram and cytokine profile of 64 patients (out of 194 patients) with multiple cytokine data points ($\geq 4$) from Wuhan Union Hospital in China. All patients were divided into mild & moderate, severe and critical groups based on their clinical symptoms, according to the 7th Version of the Novel Coronavirus Pneumonia Diagnosis and Treatment Guidance (*19*).

Here we mainly focus on the immune response features of the COVID-19 patients, which imposes huge impact on patient's clinical status. Similar to the immune efficacy $e$, we define the clinical efficacy of innate and adaptive cellular immunity of patients as $e^*(t) = Neut\%(t) \times [Monocyte\%(t) + Lymphocyte\%(t)]$, where $Neut\%(t)$, $Monocyte\%(t)$, $Lymphocyte\%(t)$ are respectively the proportion of neutrophils, monocytes and lymphocytes in peripheral blood at time point $t$.

In Fig. 3A, we classified the patients by their IL-6 level and averaged $e^*$ value ($\overline{e^*}$) into three classes. The class 1 profile (31 patients) has $\overline{e^*} > 0.15$ and IL-6$_{max}$ < 200 pg/mL, includes 15 mild patients and 16 severe patients. The class 2 profile (20 patients) has $\overline{e^*} > 0.15$ and IL-6$_{max}$ > 200 pg/mL, includes 9 mild, 10 severe and 1 critical patient. The class 3 profile (8 patients) has $\overline{e^*} < 0.15$ and IL-6$_{max}$ > 200 pg/mL, includes 1 severe and 7 critical patients. 5 patients belong to none of the above groups and are taken as exceptions.

In Fig. 3B, we aligned and averaged the time course of the three classes patients' peripheral blood cell and cytokine profile. We particularly fixed the class 2 and 3 patients' IL-6 peak at the 30[th] day and smoothed the curve using a triple-window averaging method (see this method in section 5.1 of SM). Time series data of the three classes patients differ significantly in the neutrophil, lymphocyte, IL-6, IL-10 level and immune efficacy. The class 1 patients exhibit low neutrophil counts, low IL-6 level (milder inflammatory response), high lymphocyte counts and high $e^*$ level, indicating mild symptom and effective immune response. The class 1 patients are less likely to develop into critical cases and would recover smoothly. The class 2 patients show temporarily stronger inflammatory response (higher IL-6 level) than the class 1, whose IL-6 peak usually has a 3-day full width at half maximum (FWHM). The class 3 patients show a longer period of cytokine storm status with IL-6 peak and a 13-day FWHM. In addition, the class 3 patients with lymphopenia and low $e^*$ value suggest the incompetent immune response and possible exhaustion of T cells; the significantly elevated level of neutrophils and IL-10 peak with a 7-day FWHM might indicate the occurrence of secondary bacterial infection and the emergence of MDSC, both could attribute to chronic infection and poor clinical outcome.

Thus, we correspond the class 1, 2 and 3 COVID-19 patients to the *in silico* mode 1, 2 and 3 immune response against SARS-CoV-2 infection respectively. We hope our immunity-based classification method can serve as an indicator of patient's immune response and clinical conditions.

Figure 3.

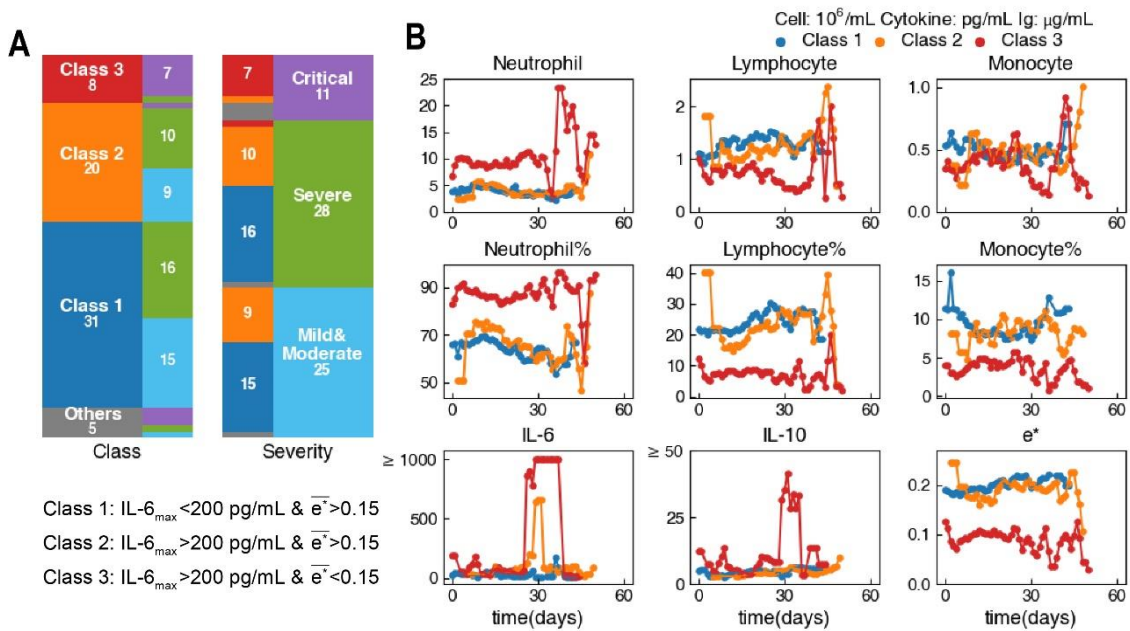

Fig. 3. The 64 clinical patients are classified by their IL-6 level and the immune efficacy in peripheral blood. (A) We identify 31 patients as the class 1 profile (15 mild & moderate patients, 16 severe patients), 20 patients as the class 2 profile (9 mild & moderate patients, 10 severe patients and 1 critical patient) and 8 patients as the class 3 profile (1 severe patient and 7 critical patients). (B) Averaged time series of the class 1, 2 and 3 patients. Compared to the steady curve of class 1 patients, class 2 and 3 patients show different extent of inflammation, characterized by the level and width of IL-6 peak. In addition, the class 3 patients show low $e^*$ level.

Combing the theoretical results and clinic classification, for the *in silico* patients with different immune response modes, we simulated their disease processes and possible clinic outcomes, and try to put forward the available treatment strategies.

The main treatment strategies of COVID-19 are to prevent the potential chronic infection and cytokine storm, increase the immune efficacy, and moderately inhibit excessive inflammatory response. Here we consider several mostly discussed agents for COVID-19: (1) Antiviral drugs (AntV), preventing viral cell entry or inhibiting the production of new progeny virus (*20*). (2) IFN-γ, increasing innate immune response and augmenting the successive adaptive immune response (*21*). (3) Monoclonal antibody (Ig), blocking viral receptor activation, suppressing the virus and modifying the inflammatory response (*22*). (4) Glucocorticoids (GC), preventing excessive immune response that causes extensive tissue damage and inhibiting cytokines from production and taking effects (*23, 24*).

We simulated the course of disease of *in silico* patients without any treatments, and examined the outcome of different treatment strategies to determine the effect of the above drugs and find reasonable combination of treatments. To quantify our results, we defined a model-based Q value to assess patient's status and examine the effects of the above medications. The scoring function Q includes patient's maximum immune efficacy $e^*_{max}$, respiratory capacity (defined by minimum healthy lung epithelial cells $[H]_{min}$), inflammation level (defined by maximum IL-6 level $[IL-6]_{max}$), and whether chronic infection happens. It is formulated as:

$$Q \equiv \left(1 + q_1 \frac{e^*_{max}}{e^*_c + e^*_{max}}\right)\left(1 + q_2 \frac{[H]_{min}}{[H]_c + [H]_{min}}\right)\left(1 + q_3 \frac{[IL-6]_c}{[IL-6]_c + [IL-6]_{max}}\right)\left(1 + q_4 \delta_{[nCoV]}\right),$$

where we set $q_1 = 0.4, q_2 = 0.4, q_3 = 3, q_4 = 1, e^*_c = 10, H_c = 30 \times 10^6 / mL$, $[IL-6]_c = 2000 pg/mL$. The Kronecker function $\delta_{[nCoV]}$ equals to 1 when final viral load is zero, equals to 0 when virus is not cleared.

We first assessed the effect of the above drugs used singly on the *in silico* patients in different modes, illustrated by the change in Q value, as in Fig. 4A. Treatment periods are divided into early (PAD 0~7), middle (PAD 7~14) and late stage (PAD 14+) referred to the estimation in (*17*). Effects of AntV, IFN-γ, Ig are positive during early stage for their role in limiting virus invasion and tissue damage, effect of GC is also positive during early stage for prevention of CRS. During middle stage, AntV and IFN-γ change little of the patients' score, but AntV is still recommended during the whole course of disease for preventing superinfection (this situation is seen in our simulation results). Usage of GC on mode 1 and 2 patients during middle stage decreases patient's Q score and is not recommended. Middle and late stage Ig benefits the most on mode 3 patients for their role in cooperating with host's immunity.

We next tested all combinations of treatment strategies with different kinds of drugs used during different stages of *in silico* patients with reasonable treatment strategies summarized in Fig. S8. Antiviral drugs are recommended during the whole course of infection. The mode 1 patients only need antiviral drugs for their recovery. For mode 2 patients, usage of IFN-γ accents innate immune efficacy and helps contain the initial tissue damage, while usage of GC can reduce inflammation in advance of cytokine storm. For mode 3 patients, cytokine storm lasts longer thus it is desired to prolong the usage of GC during early and middle stage, while during the late stage monoclonal antibodies act as backup for adaptive immunity and help clear the virus. Simulation results (ensemble averaged) for mode 2 and 3 patients with the above treatment strategies are shown in Fig. 4C and D. To assist with the formation of treatment plans, we adopted CPCA on our simulation results and identified several early stage biomarkers distinguishing the patients into mode 1, 2, and 3, including viral load, IL-6, IL-10, TNF-α and Ig, shown in Fig. S5 in SM.

Bacterial co-infection is tested positive in more than 50% of the deceased patients (*2, 25*), which we also found clinically. We simulated the course of disease with bacterial co-infection arising from mode 2 and 3, and denoted them as mode 2*, mode 3.1* and mode 3.2*, where mode 3.1* patients end up with uncleared virus, while mode 2* and mode 3.2* patients become virus-negative. Detailed assumptions, definitions and time courses concerning bacterial co-infection are in section 7 in SM.

In bacterial co-infection, antibiotics (AntBio) is also included into treatment plans. Similar to Fig. 4A, we assessed the effect of each drug in mode 2*, 3.1* and 3.2* patients, shown in Fig. 4B. Effect of early usage of GC alone turns negative due to its immunosuppressive effect, which in turn increases tissue damage and chances of bacterial co-infection, leading to extensive inflammatory response induced by bacteria. However, early and middle usage of GC, accompanied by AntBio, still makes a favorable plan, as shown in Fig. 4E.

Figure 4

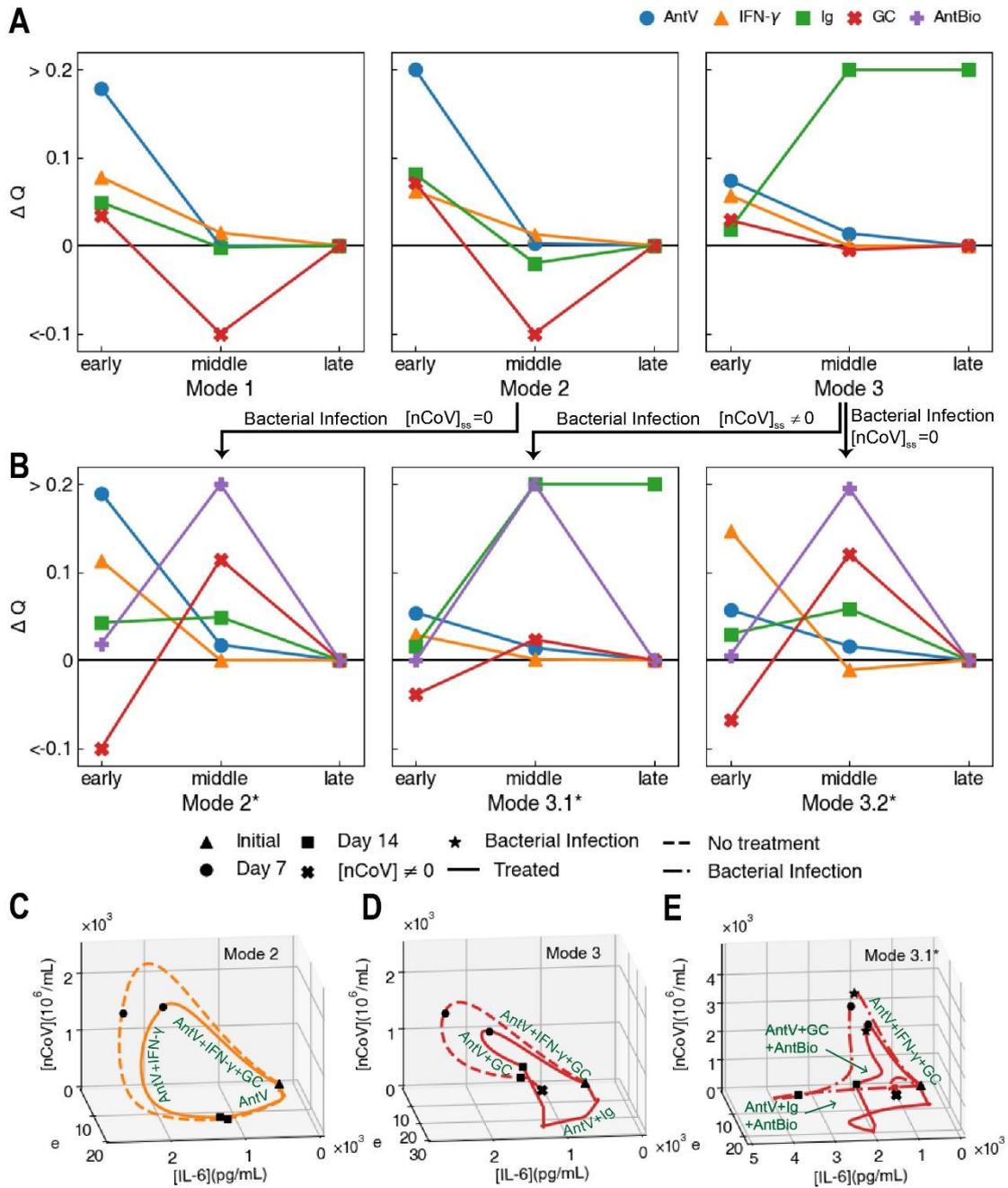

Fig. 4 Treatment strategies on *in silico* patients. (A) Effects of different drugs in improving patient's status (Q value) when used singly at different COVID-19 development stages. Median value of $\Delta Q$ in the whole sample is taken to assess the efficacy of the drugs. (B) Effects of different drugs on patients with bacterial co-infection. (C~E) The ensemble averaged immune response trajectories untreated (in dashed line) and the trajectories treated (in solid line) of the mode 2 patients (C), the mode 3 patients (D) and the mode 3.1* patients with bacterial co-infection (E).

We summarized our results in flowchart in Fig. 5, as a reference for an ideal procedure in treating COVID-19 patients.

Figure 5

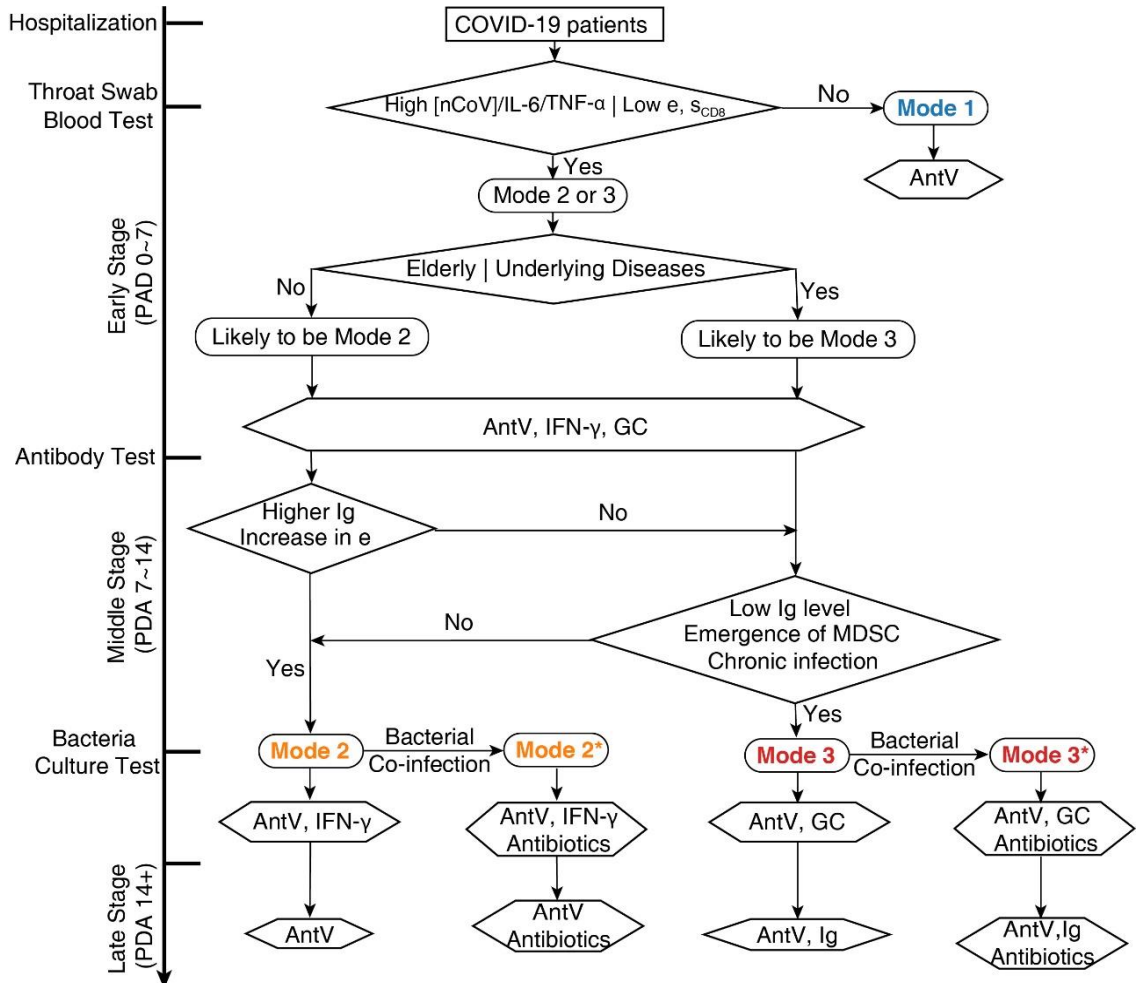

Fig. 5. Procedure for identifying the different immune modes and formulating proper therapeutic schedules.

# Discussion

The immune system is a complex defense system that protects us from the pathogens, including virus, bacteria, parasites and other invaders. The response and regulation of immune system involve dozens of cell types and hundreds of signal molecules with different ligand-receptor interactions, while these lymphocytes and molecules circulate in the whole body to clear invaders and kill infected cells. The immune system usually works quickly, effectively and resiliently. It prepares in advance through lymphocyte abundance and pool; it learns from experience by memory lymphocyte. On the other hand, it has checks and balance to prevent the over activation of immune system.

However, when a new virus or invader infect the host and there is no effective drug treatment, the host immune system might lose the balance and exhaust the lymphocyte abundance and pool. This may result in poor clinical outcome and even the epidemic spread. In the last decade of the twentieth century, during the treatment of HIV, hepatitis B virus and hepatitis C virus infection, the mathematical modeling interpreting quantitative clinical data has made a significant contribution to understand the dynamics of these viruses and the drug treatment strategy (*26-30*).

Since 2019, the ongoing of COVID-19 pandemic has sickened millions and killed more than 1 million people worldwide. There is still a lack of reliable effective treatment strategies toward the different immune status in patients. In this study, we constructed a mathematical model to describe the dynamic response of immune system. We classified the COVID-19 development processes into three typical modes of immune responses, and put forward effective treatment strategies for relevant immune response processes. In this work, we only focus on the immune response, and ignore other clinical symptoms such as multiple system and organ failure and pathological damage of other organs. We simplified the innate and adaptive immune network and chose CD4+ and CD8+ T cells to motivate and orchestrate the immune responses to SARS-CoV-2 infection. More quantitative and multiple time point clinical data are needed to verify our model and predictions, especially the treatment strategies for different modes of immune response. In our recent unpublished work, we are investigating more types of immune cells including the memory T and B cells, the bacteria infection and MDSC, together with the recirculation among blood, lung, lymph node, spleen and bone marrow. In addition, we hope our work can help to classify the patients clinically by their early period hemograms and cytokine profiles, and to choose effective treatments based on their dynamic immune status. Furthermore, we hope that our approach can be adapted to other kinds of viral and bacterial infections, and can be applied to describe and predict the cytokine storm on CAR-T immune treatment (*31, 32*).

In summary, our work provides a quantitative framework about the complex interactions between virus infection and host immune response. More feedbacks with clinical treatment and data will help us to obtain the systemic and quantitative understanding of dynamics that immune system responds to the infection of SARS-Cov-2 virus.

# Acknowledgements

The authors are grateful to Hangle Wang, Siyue Yu, Xin Gao, Dr. De Zhao for their helpful cooperation and discussions.
The authors appreciate Dr. Zhengfan Jiang, Dr. Bin Li, Dr. San Wang, Dr. Leihan Tang, Dr. Zhiyuan Li, Dr. Long Qian and Dr. Chao Tang for helpful discussions.
This work was supported by National Key R&D Program in China (Grants No. 2018YFA0900200 and 2020YFA0906900) and National Natural Science Foundation of China (Grant No. 12090051).